\documentstyle[tighten,preprint,pre,aps,psfig]{revtex}

\begin{document}

\draft

\title{ The three species monomer-monomer model: \\
A mean-field analysis and Monte Carlo study}

\author{Kevin E.\ Bassler and Dana A.\ Browne}

\address{
Department of Physics and Astronomy, Louisiana State University, Baton
Rouge, LA 70803 }

\date{October 28, 1996}

\maketitle{}

\begin{abstract}

We study the phase diagram and critical behavior of a one dimensional
three species monomer-monomer catalytic surface reaction model.  Static
Monte Carlo simulations are used to roughly map out the phase diagram
consisting of a reactive steady state bordered by three equivalent
unreactive phases where the surface is saturated with one monomer
species.  The transitions from the reactive phase are all continuous,
while the transitions between poisoned phases are first-order. Of
particular interest are the bicritical points where the reactive phase
simultaneously meets two poisoned phases.  A mean-field cluster
analysis fails to predict all of the qualitative features of the phase
diagram unless correlations up to triplets of adjacent sites are
included.  Scaling properties of the continuous transitions and the
bicritical points are studied using dynamic Monte Carlo simulations.
The transition from the reactive to a saturated phase shows directed
percolation critical behavior, while the universal behavior at the
bicritical point is in the even branching annihilating random walk
class.  The crossover from bicritical to critical behavior is also
studied.

\end{abstract}


\maketitle

\narrowtext

\section{Introduction}

Nonequilibrium models with many degrees of freedom whose dynamics
violate detailed balance arise in studies of biological populations,
chemical reactions such as heterogeneous catalysis, fluid turbulence,
and elsewhere.  The macroscopic behavior of these models can be much
richer than that of systems in thermal equilibrium, showing organized
macroscopic spatial and temporal structures like pulses or waves, and
even spatiotemporal chaos.  Even the steady state behavior can be far
more complicated, involving for example scale invariance at generic 
parameter values, and critical behavior distinct from any
equilibrium models.  However, like their equilibrium cousins,
systems at continuous transitions between nonequilibrium steady states
show universal behavior that is insensitive to microscopic details and
depends only on properties such as symmetries and conservation laws.

One place where such nonequilibrium models appear is in the study of
chemical reactions occurring on catalytic surfaces, which show a variety
of interesting behavior including nonequilibrium phase transitions,
temporal oscillations, spiral waves, and chemical chaos\cite{bookref}.  
In order to
help understand these complicated processes, a number of simple models
have recently been proposed that attempt to capture the essential
physics\cite{M+D}.

Ziff, Gulari, and Barshad (ZGB) proposed a monomer-dimer reaction model
to explain some features of CO oxidation on a noble metal surface \cite{ZGB1}.  
In their model, monomers representing CO molecules and dimers representing
O$_2$ molecules adsorb on a lattice. Immediately upon adsorption, the
O$_2$ dimers dissociate into two O monomers. CO monomers and O monomers
occupying nearest neighbor sites then react to form a CO$_2$ molecule
that immediately desorbs, leaving two vacant lattice sites. In the
limit of infinitely fast reactions (the adsorption controlled limit),
where the only parameter of the model is the relative adsorption rate
of CO molecules $y_{CO}$, they found in two dimensions that
there are three phases: An O$_2$, or dimer poisoned state for $y_{CO} <
y_1$, a CO, or monomer poisoned phase for $y_{CO} > y_2$, and a
reactive phase for $y_1 < y_{CO} < y_2$. At $y_1$ the fraction of each
species changes continuously, indicating that the dimer poisoning
transition is continuous.  At $y_2$ the monomer poisoning transition is
first order, with the densities of the different species changing
discontinuously.  In one dimension, the ZGB monomer-dimer reaction model
has no reactive phase, only monomer-poisoned and dimer-poisoned phases
separated by a first-order transition\cite{ZGB1d}.

An even simpler catalytic reaction model can be constructed by
replacing the dimer species in the ZGB model with a second monomer
species. This monomer-monomer model has a long history\cite{MM}, 
and in fact certain analytic results for this model have been obtained
in the reaction-controlled limit of the model\cite{MMar}. In this
model two different monomer species, call them A and B, adsorb on a
lattice where nearest neighbor AB pairs react and an AB molecule
desorbs. However, the phase diagram for this model does not contain a
reactive steady state in any number of dimensions,  neither in the
adsorption controlled nor in the reaction controlled limit.  The phase
diagram consists only of A and B poisoned states, and a first-order
transition between them.

The dimer poisoning transition in the ZGB model is one of the most
common types of continuous phase transitions in nonequilibrium models.
It is a transition to a single absorbing, noiseless, steady
state, the term absorbing indicating the state cannot be left
once it is reached.  Other examples include directed percolation
(DP)\cite{DP1,DP2}, the contact process\cite{CP}, 
auto-catalytic reaction models\cite{ABW}, 
and branching annihilating random walks with
odd numbers of offspring\cite{BAW2,BAW3}.  Both renormalization group
calculations\cite{DP1,RFT} and Monte Carlo
simulations\cite{DP2,CP,ABW,BAW2,BAW3,ZGB2} show that these models form
a single universality class for a purely nonequilibrium model with no
internal symmetry in the order parameter.

Recently, a number of models with continuous adsorbing transitions in a
universality class distinct from directed percolation have been
studied.  These models include probabilistic cellular automata
models studied by Grassberger {\em et al.\/}\cite{PCA}, certain kinetic
Ising models \cite{NKI}, the interacting monomer-dimer
model\cite{IMD1,IMD2}, and branching annihilating random walks with an
even number of offspring (BAWe)\cite{BAW2,BAWe}.  All of these models
except for the BAWe have two equivalent absorbing states indicating the
importance of symmetry of the adsorbing state to the universality
class.  However, the universal behavior of this new class is apparently
controlled by a dynamical conservation law.  If the important
dynamical variables in this class are defects represented by the
walkers in the BAWe model and the walls between different saturated
domains in the other models,  the models have a ``defect parity''
conservation law\cite{PCA} where the number of defects is conserved modulo 2.
Recent field theoretic work confirms this viewpoint\cite{CardyRG}.

In a recent Letter\cite{letter}, we introduced a monomer-monomer
reaction model with three different monomer species. This model could
represent either a system with three different chemical species or an
auto-catalytic reaction system in which one chemical species can adsorb
on three different types of surface sites.  Using static and dynamic
Monte Carlo simulations, we determined the phase diagram and studied
the phase transitions in the one dimensional version of the model, and
showed that it has continuous adsorbing transitions to both one and two
equivalent noiseless states.  It is therefore is a good model to study
the role of symmetry in adsorbing phase transitions.

In this paper we expand those results, providing more details of our
simulation methods and of the results, again restricting our
consideration the the one-dimensional version of the model. We also
include a mean-field cluster analysis of the model including up to
triplets of adjacent sites. The paper is organized as follows. In the
next section we define the model and show the phase diagram of the
model, as determined by simulations. The following section presents the
mean-field analysis. Section IV contains the details and results a
detailed Monte Carlo study of the dynamic scaling behavior at the
various phase transitions, and of the crossover behavior between the
different types of scaling behavior. In the last section we summarize
our results.

\section{The model}

\begin{figure}
\vspace{1in}
\centerline{\psfig{file=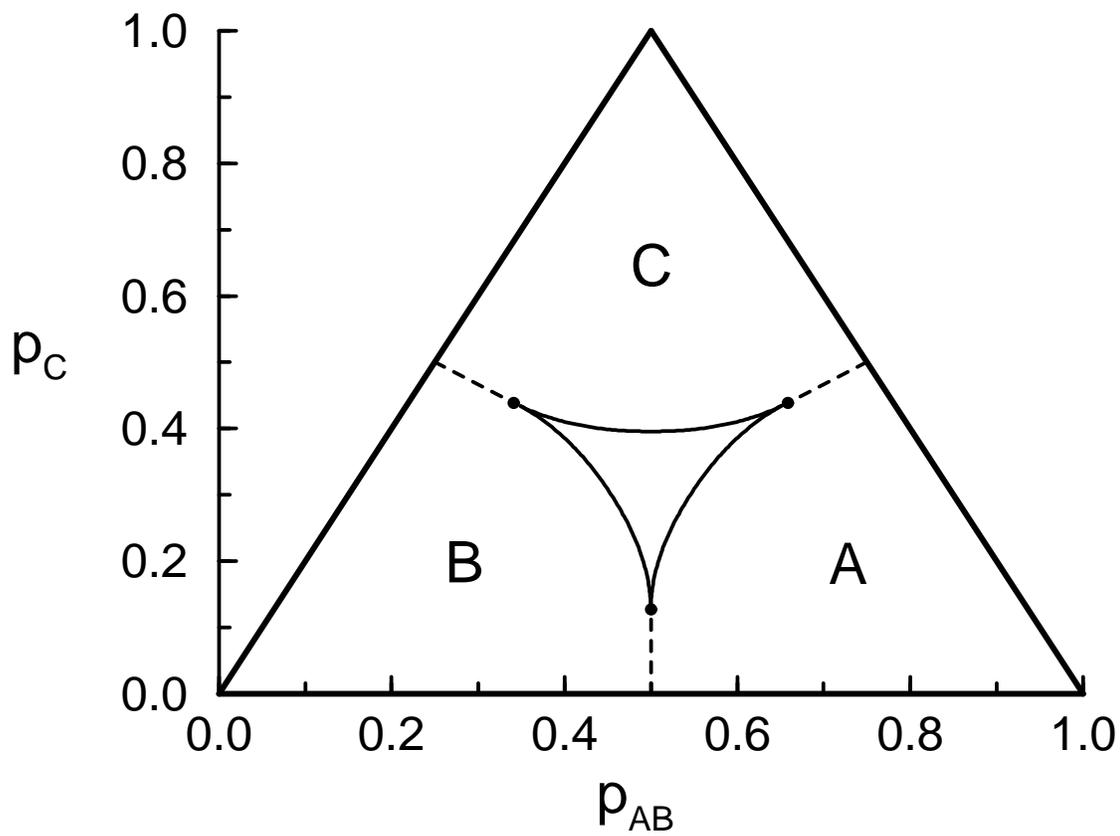,width=6.0in}}
\vspace{0.5in}
\caption{ Phase diagram showing three saturated phases
(indicated by the letters), and a reactive phase
(the unlabeled center region).  Solid lines indicate continuous transitions.
Dashed lines indicate first-order transitions.  Bicritical points (filled
circles) occur
where two critical lines meet a first-order line. }
\label{fig1}
\end{figure}

Our three species monomer-monomer model is defined by two fundamental
dynamic processes:  (a) monomer adsorption at sites of a substrate,
and (b) the annihilation reaction of two dissimilar monomers adsorbed
on nearest-neighbor sites of the substrate.  Here we consider the model
only in the adsorption controlled limit where process (b) occurs
instantaneously.  Calling the monomer species $A$, $B$ and $C$, the
parameters in the model are then the relative adsorption rates of the
different monomer species $p_A$, $p_B$, and $p_C$, such that
$p_A+p_B+p_C=1$.  Using static Monte Carlo simulations to get the rough
picture, and refining it with dynamical Monte Carlo studies described
below, we find the ternary phase diagram for the model is shown in
Fig.~\ref{fig1}.  In this figure, the horizontal axis corresponds to
the relative adsorption rate of $A$ and $B$ monomers
$p_{AB}=p_A/(p_A+p_B)$.  The absorbing phases, where one monomer
species saturates the chain, occupy the corners of the phase diagram.
In the center of the phase diagram is a
a reactive steady state.  There are continuous phase
transitions from the reactive phase to the saturated phases, but the
monomer densities undergo discontinuous, first-order, transitions from
one saturated state to another.  The points where the reactive phase
and two saturated phases meet are {\em bicritical} points\cite{BCP}
where two lines of continuous transitions meet a line of first-order
transitions.

\section{Mean-field theory}

To analyze the kinetics of the three species monomer-monomer model, it
is useful to perform a mean-field analysis.  While such analysis
neglects long-range correlations and thus cannot be expected to
properly predict critical properties, it should properly predict the
qualitative structure of the phase diagram, including the existence of
continuous transitions and multi-critical points.  The mean-field
analysis also provides a starting point for studying the importance of
such fluctuations, which, of course, become particularly important near
continuous phase transitions.  The mean-field approach we
use\cite{DickmanMFT} studies the time evolution of clusters of sites,
the approximation coming in truncating the probabilities of observing
clusters of larger size into probabilities for smaller size clusters.
The simplest form is the site approximation where probabilities of
observing certain nearest neighbor pairs is replaced by the produce of
the average site densities.  Better approximations can be obtained
systematically by replacing the actual configuration of larger
clusters, i.e. pairs, then triplets, and so on, with the average
density of those clusters.  The analysis presented below includes
clusters consisting of up to triplets of adjacent sites.

\subsection{Site approximation}

At a particular time, a lattice with $N$ sites will have $N_V$
vacancies, the remaining sites being filled with $N_A$,
$N_B$, and $N_C$ numbers of $A$, $B$, and $C$ monomers respectively.
The density of $A$ monomers is 
$x_A \equiv (N_A / N)$, with corresponding definitions for $B$, $C$ and
vacant ($V$) sites.  We have the obvious constraint
\begin{equation}
x_V + x_A + x_B + x_C \equiv 1.
\label{siteconstraint}
\end{equation}
In the site approximation all correlations are neglected, so
that $x_V^2$ is the probability that a given pair of lattice sites
are occupied by two vacancies. The rate equations for the $A$ monomer density is
\begin{equation}
{d x_A \over dt}
 =  p_A x_V (1 - x_B - x_C )^2
 - (p_B + p_C) x_V \left[ 1 - (1 - x_A )^{2} \right]
\label{rateeq}
\end{equation}
with similar equations for $x_B$ and $x_C$. 
The first term on the right hand side
of Eq.~(\ref{rateeq}) is the rate of $A$ monomer adsorption multiplied
by the probability that an adsorbing $A$ monomer will find a vacant
site that has no
$B$ or $C$ monomers adsorbed on adjacent sites.
The second term is the rate at which $B$ or $C$ monomers find
a vacant site with at least one adjacent adsorbed $A$ monomer 
to react with.

Equations~(\ref{rateeq}) have steady state solutions
corresponding to each of the three poisoned states, as well as
one corresponding to the reactive steady state.
To find the site approximation phase diagram, we analyzed the
stability of those solutions as a function of the rates $\{p_\alpha\}$
by examining the eigenvalues of the Jacobian matrix
for linearized rate equations.

\begin{figure}
\vspace{1in}
\centerline{\psfig{file=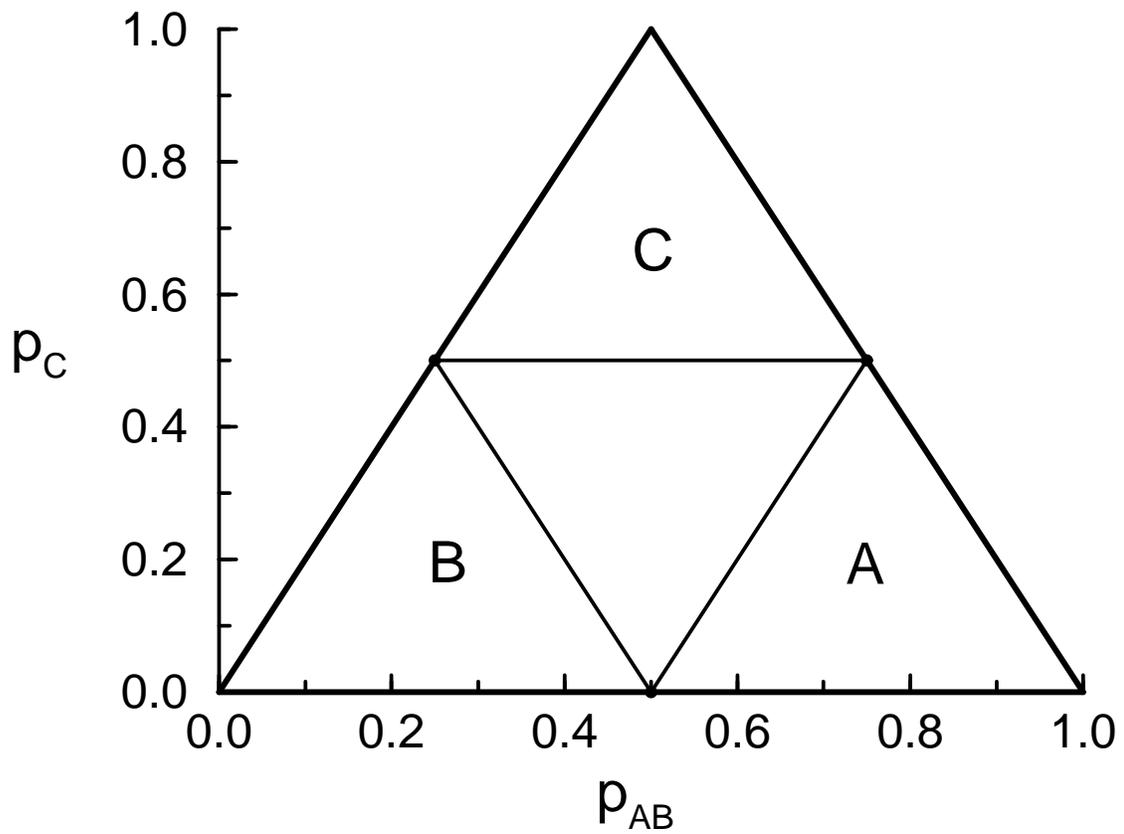,width=6.0in}}
\vspace{0.5in}
\caption{ Phase diagram in the site approximation.
Transitions between the reactive phase (unlabeled) and the three
saturated phases (indicated by the letters) are continuous.
Note that unlike the actual phase diagram shown in Fig.~\protect{\ref{fig1}}
the continuous
transition lines are straight and the bicritical points where two
continuous transition lines meet are on the edge of the phase diagram.
}
\label{sitepd}
\end{figure}

For example, the Jacobian matrix for the $A$ poisoned state has two
zero eigenvalues and one eigenvalue of $p_B + p_C - p_A = (1-2p_A)$.
This third eigenvalue shows that the $A$ poisoned state is stable only
for $p_A > 1/2$.  Corresponding results hold for the other poisoned
states, leading to the site approximation phase diagram
shown in Fig.~\ref{sitepd}.

As the phase boundaries are approached from the reactive phase, the
monomer densities vanish continuously, indicating a continuous
transition to an absorbing state.  The points on the edge of the phase
diagram where two different poisoned phases meet the reactive phase are
bicritical points.

\subsection{Pair approximation}

We improve the site approximation by properly accounting for the
correlation of nearest neighbor pairs and approximating the correlations
of triples and larger clusters.
We define $N_{ij}$ as the number of bonds connecting nearest neighbor sites
occupied by $i$ and $j$ monomers ($A$, $B$, $C$, or $V$), where the monomer
$i$ occupies the site to the left of monomer $j$, and we have $N_{ij}=N_{ji}$.  
Since we are studying
one dimension, the number of bonds equals the number of sites $N$,
so the bond densities are
defined by
$$
x_{i j} \equiv {N_{i j} + N_{ji}\over N}, \qquad \qquad i\neq j
$$
and
$$
x_{ii}\equiv {N_{ii}\over N}
$$
There are seven different allowed types of bonds: V-V, A-A, B-B, C-C,
A-V, B-V, and C-V.  Other types of bonds, A-B, A-C, and B-C, are
forbidden in the adsorption controlled limit we are considering.  
The densities satisfy the constraint
\begin{equation}
x_{VV} + x_{AA} + x_{BB} + x_{CC} + x_{AV} + x_{BV} + x_{CV} = 1,
\label{pairconstraint}
\end{equation}
so only six of the $x_{ij}$ are independent.
The $A$ monomer density is given by $x_A = x_{AA} + {1\over2}x_{AV}$,
with similar expressions for the $B$ and $C$ densities.

To determine the equations of motion of the pair densities it is useful
to distinguish between the different types of events that change the
configuration.  For example, if an $A$ monomer attempts to occupy a site,
it can (1) stick, (2) react with a $B$ or (3) react with a $C$,
which we indicate respectively with the shorthand :
(1) $A\downarrow$, (2) $A\downarrow \; AB\uparrow$, and  
(3) $A\downarrow \; AC\uparrow$.  
The rate equations can be written as
$$
{d x_{ij} \over dt} = \sum_{\alpha} \Delta x_{ij}^{(\alpha)}
$$
where $\alpha$ refers to the event type, and
$\Delta x_{ij}^{(\alpha)}$ is the change in $ij$ bond density
arising from an event of type $\alpha$.

To find the different bond density changes
note that the probability $P(i|j)$ for a site to be occupied
by a monomer (or vacancy) of type
$i$, given that one of its nearest neighbors is of type $j$, is
$$
P(i|j) = {N_{ij}\over N_i} = {x_{ij} \over 2 x_{j}}
$$
for $i\neq j$, and
$$
P(i|i) = {x_{ii} \over x_{i}}
$$

The various $\Delta x_{ij}^{(\alpha)}$ are given in Table 1,
where
$$
z_{iV} = P(i|V)+P(V|V) = {x_{VV} + {1 \over 2}x_{iV} \over x_{V}}
$$
is the probability that the site to the left of a vacant site is
occupied by either an $i$ type monomer or a $V$.
The density changes due to the other event types are found by permutation.

Thus, the rate equations are

\widetext

\begin{eqnarray}
{d x_{AA} \over dt}
& = &
p_{A} \; x_{AV} \; z_{AV}
- {x_{AA} x_{AV} \over 2 x_{A}}
\left[ p_{B}(1 + z_{BV}) + p_{C} (1 + z_{CV})\right]
\nonumber \\
{d x_{AV} \over dt}
& = &
p_{A} \; z_{AV} (2x_{VV} - x_{AV})
- {x_{AV}^2\over 2x_A}
\left[ p_{B}(1 + z_{BV}) + p_{C} (1 + z_{CV})\right]
\label{bondrateeqn}
\end{eqnarray}

\narrowtext

The other equations can be found by permutation, except for
$x_{VV}$ which can be found using Eq.~(\ref{pairconstraint}).

\begin{figure}
\vspace{1in}
\centerline{\psfig{file=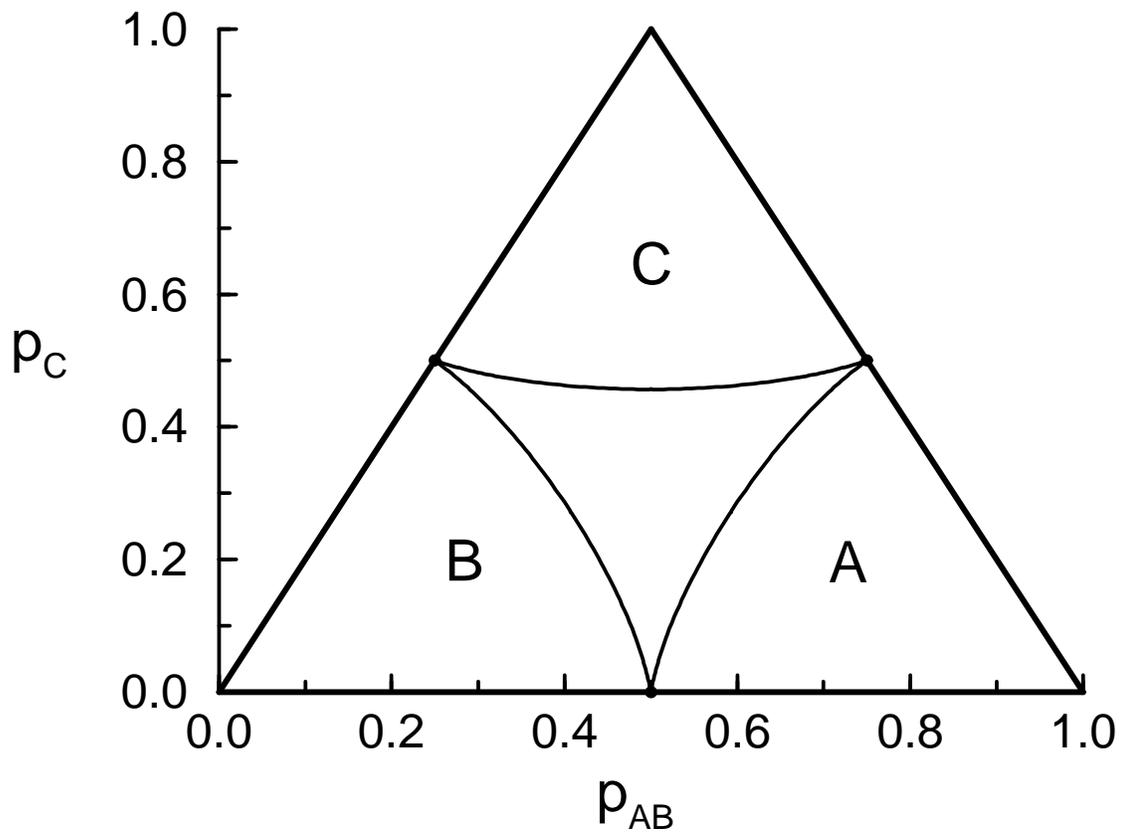,width=6.0in}}
\vspace{0.5in}
\caption{ Phase diagram in the pair approximation.
Transitions between the reactive phase (unlabeled) and the three
saturated phases (indicated by the letters) are continuous.
Note that unlike the actual phase diagram shown in Fig.~\protect{\ref{fig1}}
the bicritical points are still on the edge of the phase diagram.
}
\label{pairpd}
\end{figure}

Multiple steady state solutions to the set of six coupled bond density
rate Eqs.~(\ref{bondrateeqn}) correspond
to the reactive state (which can be found numerically), as well the
three poisoned states.  In principle, to find the phase diagram a
stability analysis of those steady state solutions could be performed.
However, we instead simply solved the six equations numerically as a
function of the parameters $p_{AB}$ and $p_C$, and looked for the
transitions to the poisoned states.  The results are shown in
Fig.~\ref{pairpd}.  The densities of the different monomer species
still change continuously as the phase boundaries are approached,
indicating that the transitions are continuous.  While the
phase boundaries are now curved as they are in the actual phase
diagram, the bicritical points are still on the edge of the
phase diagram unlike the
actual phase diagram.

\subsection{Triplet approximation}

\begin{figure}
\vspace{1in}
\centerline{\psfig{file=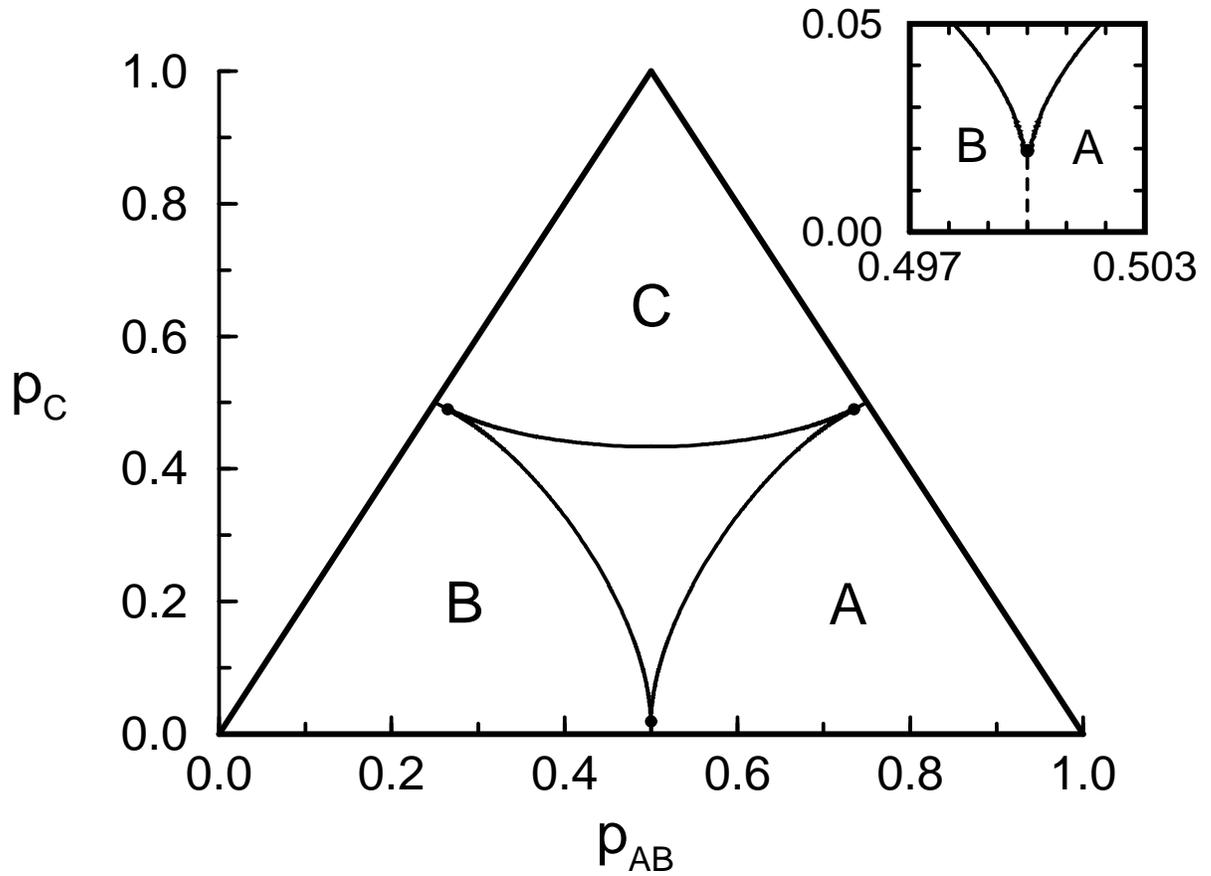,width=6.0in}}
\vspace{0.5in}
\caption{ Phase diagram in the triple approximation.
Transitions between the reactive phase and the three
saturated phases are continuous, while transitions between
saturated phases are first-order.
Inset shows a closeup of
the phase diagram near the bicritical point at the end of the first-order
line separating the A and B saturated phases.
All of the qualitative features of 
the actual phase diagram are reproduced in this approximation. }
\label{triplepd}
\end{figure}

The mean-field theory can be refined even further by considering larger
clusters. However, this systematic process rapidly increases in
difficulty.  But since even the pair approximation failed to predict
that the bicritical points occur on the interior of the phase diagram,
we pushed the cluster expansion one step further and analyzed the model
in the triplet approximation. In this approximation, clusters of three
adjacent sites are considered, thereby including the effects of
correlations up to that level.  The details of the calculation are
presented in the Appendix, but here we summarize the results.  In one
dimension, there are 19 different allowed triplets. However, 4
different constraints reduce the number of independent triplets to 15.
Numerically solving the rate equations for the densities of those 15
different triplets simultaneously, we find solutions corresponding to
the reactive steady state, as well as the poisoned states.  The phase
diagram, calculated as for pair approximation, is shown in
Fig.~\ref{triplepd}.  Finally at this level of approximation all of the
qualitative features of the actual phase diagram are predicted.  In
particular, the bicritical points appear on the interior of the phase
diagram and there are first order lines between the poisoned phases.
However, note that the size of the poisoned phases is still
underestimated by the mean-field cluster analysis, even in the triplet
approximation.  For example, the bicritical point on the $p_{AB}=0.5$
line occurs at about $p_C=0.02$ in the triplet approximation, whereas
in actuality it occurs at about $p_C=0.12$.  This indicates that
fluctuations, which are still not fully accounted for in mean-field
theory, stabilize the poisoned phases.

\section{Simulations}

To further investigate the three species monomer-monomer model we also
used time-dependent Monte Carlo simulations.  Particularly useful for
studying critical properties, the method is a form of ``epidemic''
analysis\cite{BAW3,BAWe,DMCS} in which the average time evolution of a
particular configuration that is very close to an adsorbing state
(defect dynamics), or very close to a minimal width interface between
two different adsorbing states (interface dynamics), is measured by
simulating a large number of independent realizations.  Using this
technique we determined the universality classes of the critical and
bicritical points, and studied the critical dynamics of interfaces
between the two symmetric saturated states at the bicritical points,
and the crossover from bicritical to critical behavior, including
measuring the crossover exponent $\phi$, as well as the subcritical
behavior at the first-order lines.

Because monomers can adsorb only at vacant sites, and the total
number of vacancies on the lattice is usually very small, instead of
randomly picking a site to attempt to adsorb on, it is much more
efficient to use a variable time algorithm in which the adsorption site
is randomly picked from a list of vacant sites.  The species of monomer
chosen for adsorption is then randomly picked according to the relative
adsorption rates $\{p_{\alpha}\}$, and the time length of a step is
$1/n_V(t)$ where $n_V(t)$ is the total number of vacancies at that
time.  Thus, on average there is one attempted adsorption per lattice
site per unit time. We always start with a lattice big enough that the
active region will never reach the boundaries; it is effectively an
infinite lattice.

\begin{figure}
\vspace{1in}
\centerline{\psfig{file=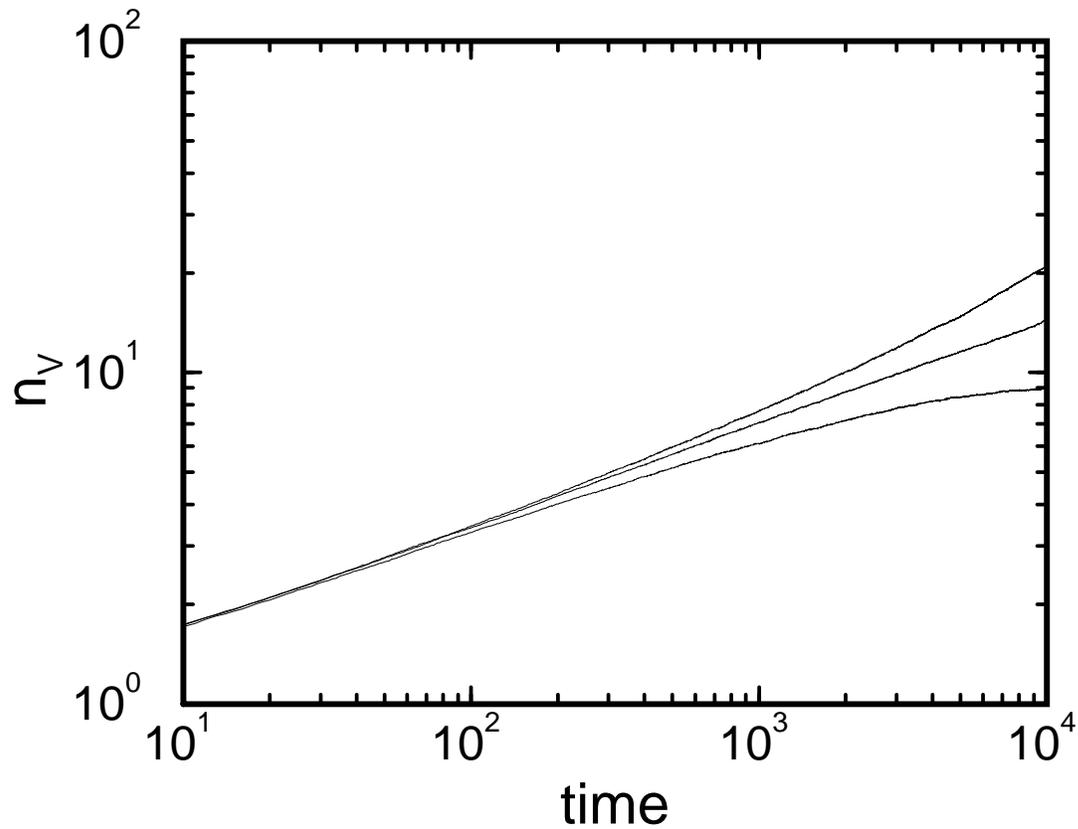,width=6.0in}}
\vspace{0.5in}
\caption{Log-log plot showing the average number of vacancies, $n_V$
as a function of time 
near the transition from the reactive phase to the C saturated phase
at $p_{AB}=0.5$.  From top to bottom,
the 3 curves correspond to
$p_C = 0.395$, 0.39575, and 0.3965.  The middle curve
corresponds to the critical point.  Note that the critical line is straight
while the other lines have curvature.}
\label{loglogplots}
\end{figure}

During the simulations we measured the survival probability $P(t)$,
defined as the probability that the system had not poisoned by
time $t$, the average number of vacancies per run
$\langle{n}_V(t)\rangle$, and the average mean-square size of the
active region per surviving run $\langle{R^2}(t)\rangle$.
At a continuous phase transition as
$t\rightarrow\infty$ these dynamical quantities obey power law behavior
\begin{equation}
P(t) \sim t^{-\delta} , \qquad
\langle{n}_V(t)\rangle \sim t^{\eta} , \qquad
\langle{R^2}(t)\rangle \sim t^{z} .
\end{equation}
Plots of the logarithms of these quantities versus the logarithm of the
time, such those shown in Fig.~\ref{loglogplots} yield
a straight line at the phase transition, and show curvature away from
the transition.

Precise estimates of the location of the
critical point and of the exponents can be made by examining the local
slopes of the curves on a log-log plot.
The effective exponent $\delta(t)$ is defined as
\begin{equation}
-\delta(t) = \{ \ln\left[ P(t)/P(t/b) \right] / \ln b \}\,,
\label{localslopes}
\end{equation}
with similar expressions for $\eta(t)$ and $z(t)$.  At the critical point,
a graph of the
local slope versus $t^{-1}$ should extrapolate to the critical exponent,
with a correction that is expected\cite{ctsnote} to be linear in $t^{-1}$.
Away from the critical point, the local slope curve should show strong
curvature away from the critical point value as $t^{-1}\to0$.

\subsection{Critical dynamics}

Figure~\ref{loglogplots} shows the data for the three dynamic
quantities near the phase transition to the $C$ saturated phase at
$p_{AB}=0.5$ plotted against time on a log-log scale.  This data was
calculated from $10^5$ independent runs of up to $10^4$ time steps at
each parameter value.  As expected, right at the critical point the
line is straight, indicating power law scaling, and away from the
critical point the lines show curvature.

\begin{figure}
\vspace{1in}
\centerline{\psfig{file=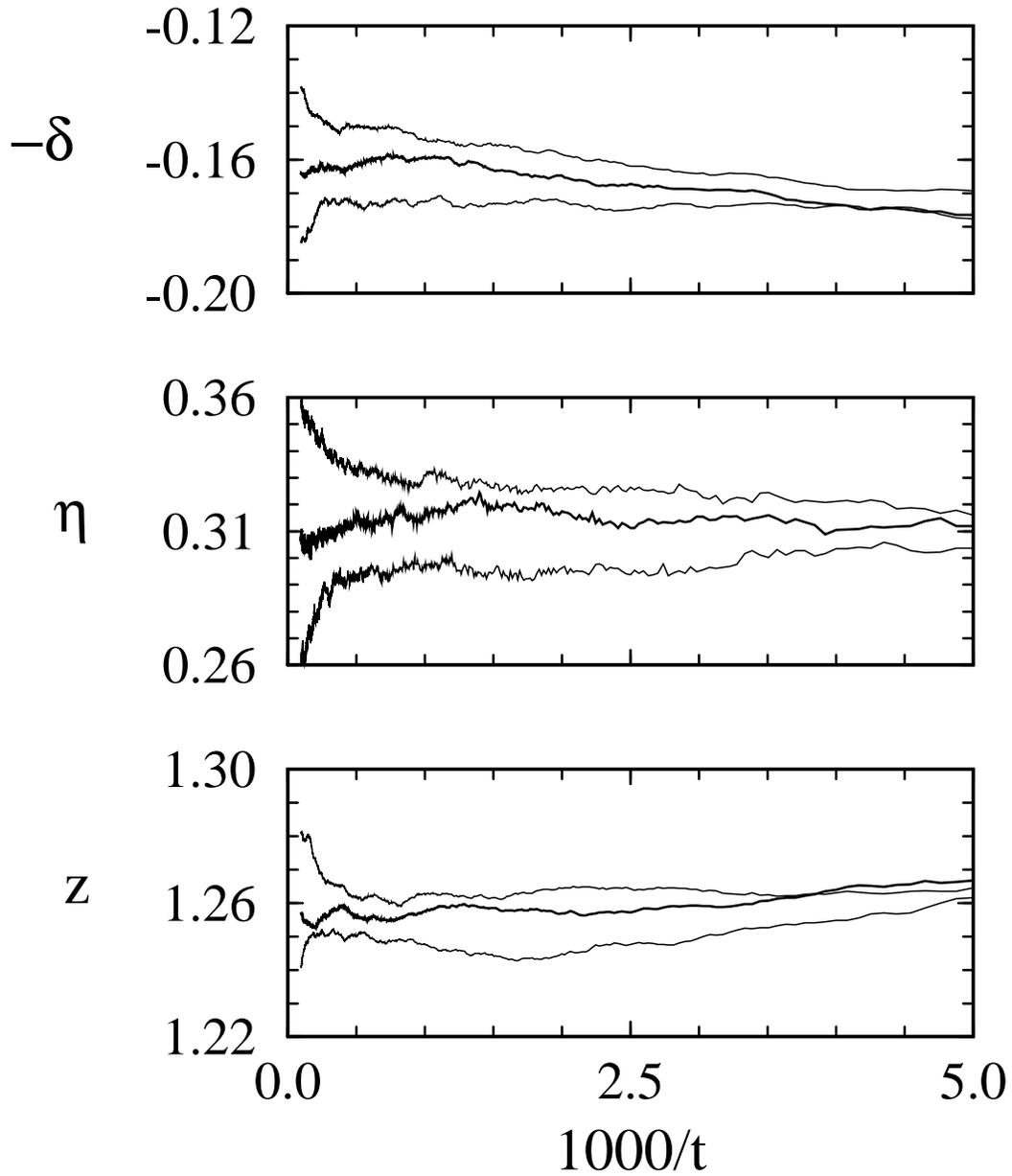,width=6.0in}}
\vspace{0.5in}
\caption{
Effective exponents using Eq.~(\protect\ref{localslopes}) with
$b=5$ for the defect dynamics
near the critical point at $p_{AB}=0.5$ on the line where the
C poisoned phase meets the reactive phase.  From top to bottom,
the 3 curves in each panel correspond to
$p_C = 0.3955$, 0.39575, and 0.3960, with the middle curve
corresponding to the critical point.}
\label{figdp}
\end{figure}

The exponents and the location of the critical point are easily and
precisely determined by taking the local slopes of this data, which are
shown in Fig.~\ref{figdp}.  We find a critical $C$ monomer adsorption
rate of $\tilde{p}_C = 0.39575(10)$, and that the critical exponents
are $\delta = 0.16(1)$, $\eta = 0.31(1)$, and $z = 1.255(15)$.  These
values are consistent with our expectation that this transition should
be in the DP universality class, for which the exponents are $\delta =
0.1596(4)$, $\eta = 0.3137(10)$, and $z = 1.2660(14)$\cite{DPexp}.  We
found similar exponents for the adsorbing transition at a number of
other points along the lines separating the reactive phase and the
saturated states (see the discussion of the crossover from bicritical
to critical behavior below), indicating the transition between the
reactive phase and any single saturated phase is always in the DP
universality class.

\subsection{Bicritical defect dynamics}

\begin{figure}
\vspace{1in}
\centerline{\psfig{file=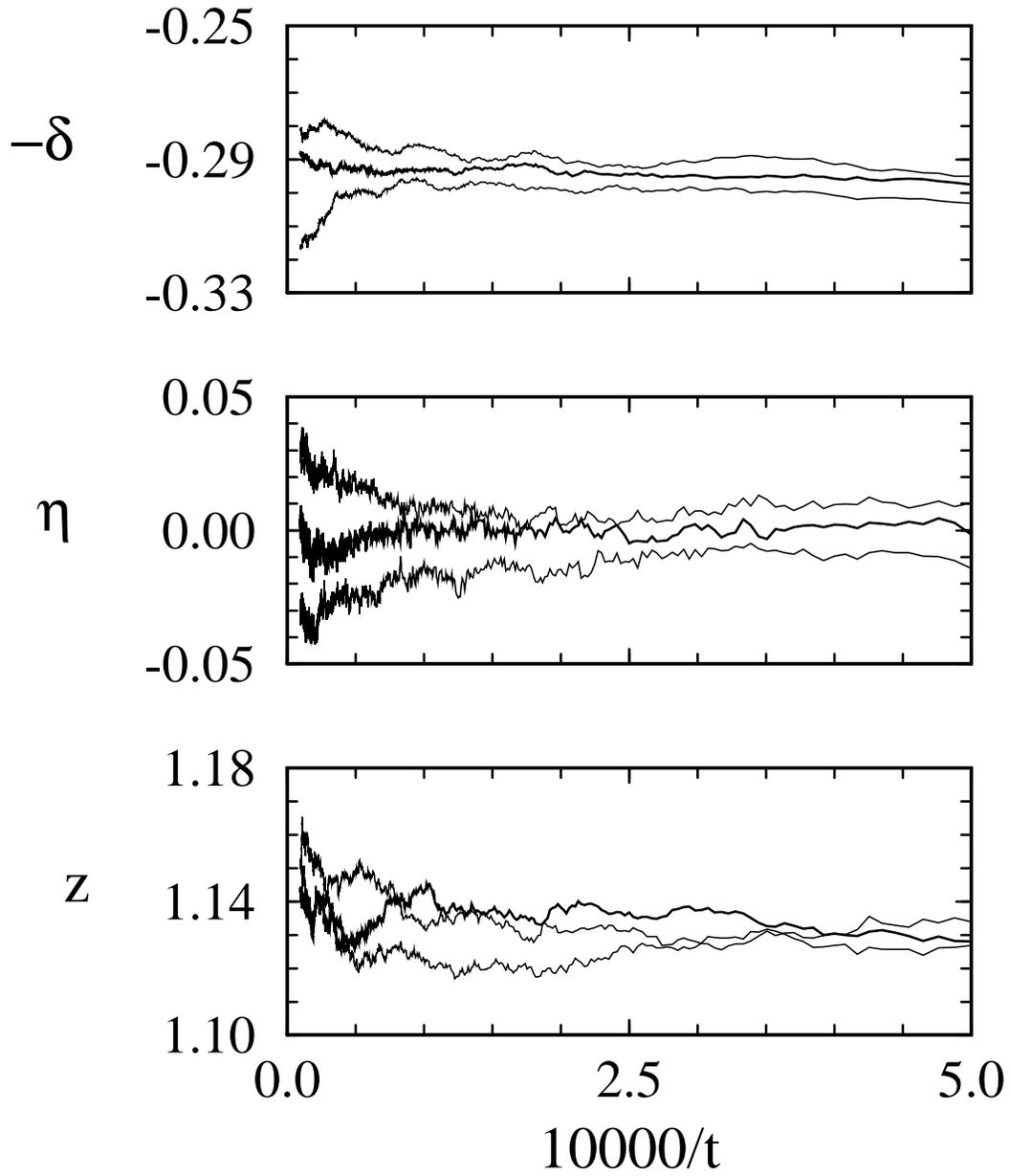,width=6.0in}}
\vspace{0.5in}
\caption{
Effective exponents using Eq.~(\protect\ref{localslopes}) with
$b=5$ for the defect dynamics
near the bicritical point where the
A and B poisoned phases meet the reactive phase
as defined in Eq.~(\protect\ref{localslopes}) with $b=5$.  From
bottom to top, the 3 curves in each panel correspond to
$p_C = 0.121$, 0.122, and 0.123, with the middle line
corresponding to the bicritical point.}
\label{fig7}
\end{figure}

The same kind of analysis at the bicritical point at $p_{AB}=0.5$,
using an initial condition of a vacancy in an A-saturated phase, yields
a bicritical point at $p_C = p_C^{*} = 0.122(1)$.   The exponents
at the bicritical point are very different, which we expect
given the presence of two-symmetry equivalent saturated
phases.  From $5\times10^5$ runs of up to $10^5$ time steps we found
the local slope data shown in Fig.~\ref{fig7}, yielding values of
$\delta= 0.29(1)$, $\eta = 0.00(1)$, and $z = 1.150(15)$.  These values
indicate that the bicritical behavior falls in the BAWe universality
class, for which $\delta = 0.285(2)$, $\eta = 0.000(1)$, and $z =
1.141(2)$\cite{BAWe}.

\begin{figure}
\vspace{1in}
\centerline{\psfig{file=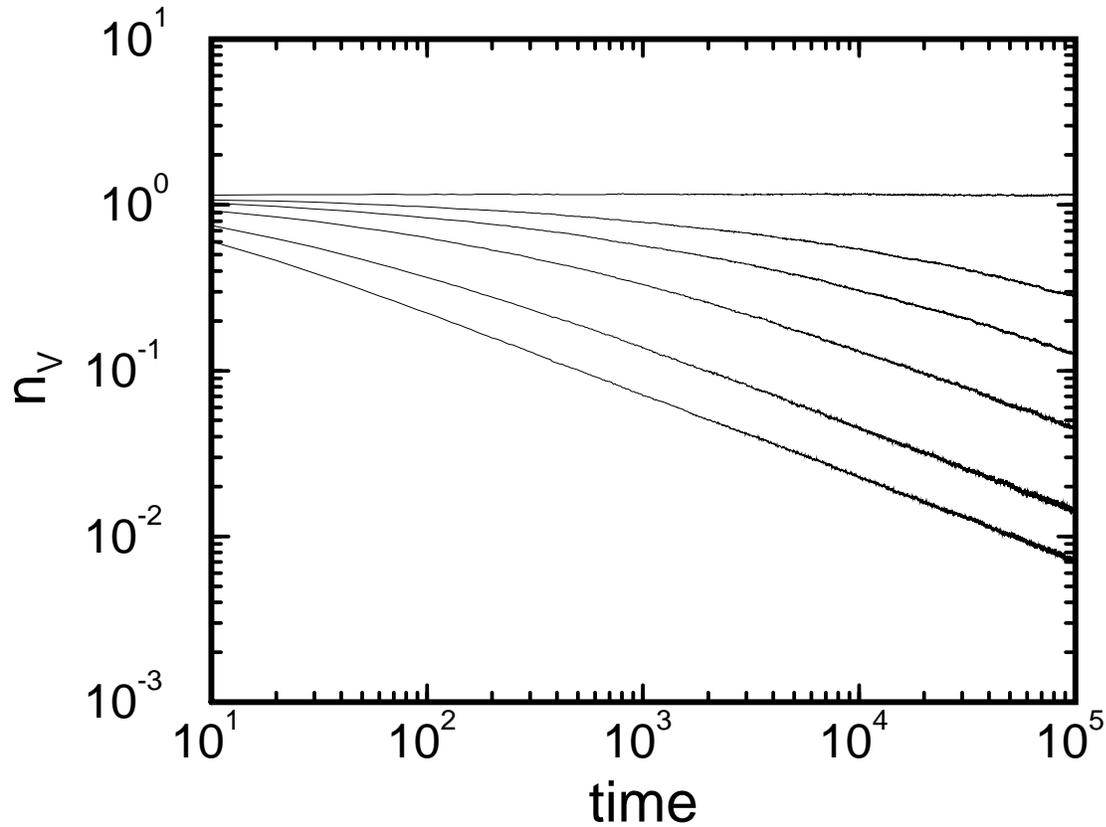,width=6.0in}}
\vspace{0.5in}
\caption{
Average number of vacancies for $p_{AB}=0.5$, and $p_C \leq p^*_C$
showing crossover from bicritical behavior to sub-critical behavior.
From bottom to top, the curves correspond to
$p_C = 0.0$, 0.04, 0.08, 0.10, 0.11, and 0.122.  The top curve ($p_C=0.122$)
corresponds to the bicritical point.  All other curves tend toward
a slope of $-0.5$ at large t.}
\label{subcritxover}
\end{figure}

For $p_C < p^{*}_C$ along the A-B coexistence line, a similar analysis
shows a crossover from the bicritical behavior to sub-critical behavior
corresponding to the well known problem of the $T=0$
one-dimensional kinetic Ising model for which dynamic exponents
$\delta = 0.5$, $\eta = -0.5$, and $z = 1$
are known exactly\cite{zeroTIsing}.
The two species version of our model, which occurs for $p_C=0$ on the edge
of the phase diagram, can be mapped onto this
kinetic Ising Model.
However, as can be seen in Fig.~\ref{subcritxover}, for $0 < p_C < p^{*}_C$
at short times the dynamic critical behavior tends to act more like the
bicritical behavior before changing to kinetic Ising model behavior at 
long times. The time which this crossover occurs increases as $p_C$ approaches $p^{*}_C$, but for all $p_C < p^{*}_C$ the long time dynamical
critical behavior corresponds to the kinetic Ising model.

\subsection{Bicritical interface dynamics}

To further analyze the importance of competition in the growth of two
equivalent saturated phases at the bicritical point we also studied the
dynamics of an interface between those two phases.  Starting with a
single vacancy between the two domains, we used two different methods
to analyze the behavior of the interface.  Since there must always be
at least one vacancy between two different saturated phases, in the
first method we ignore the survival probability $P(t)$ and take
$\delta\equiv0$.  We then measure the number of vacancies in the
interface $\langle n(t)\rangle \propto t^{\eta}$ and average size of
the interface $\langle R^2(t)\rangle \propto t^{z}$.  From
$5\times10^4$ independent runs at the bicritical point, each lasting
$10^5$ time steps, we found the other exponents to be $\eta=0.285(10)$
and $z=1.14(2)$.  This type of interface dynamics has been used to
study the properties of critical interfaces in other models in the BAWe
class, where similar results for $\eta$ and $z$ were
obtained\cite{IMD2,BAWe}.

\begin{figure}
\vspace{1in}
\centerline{\psfig{file=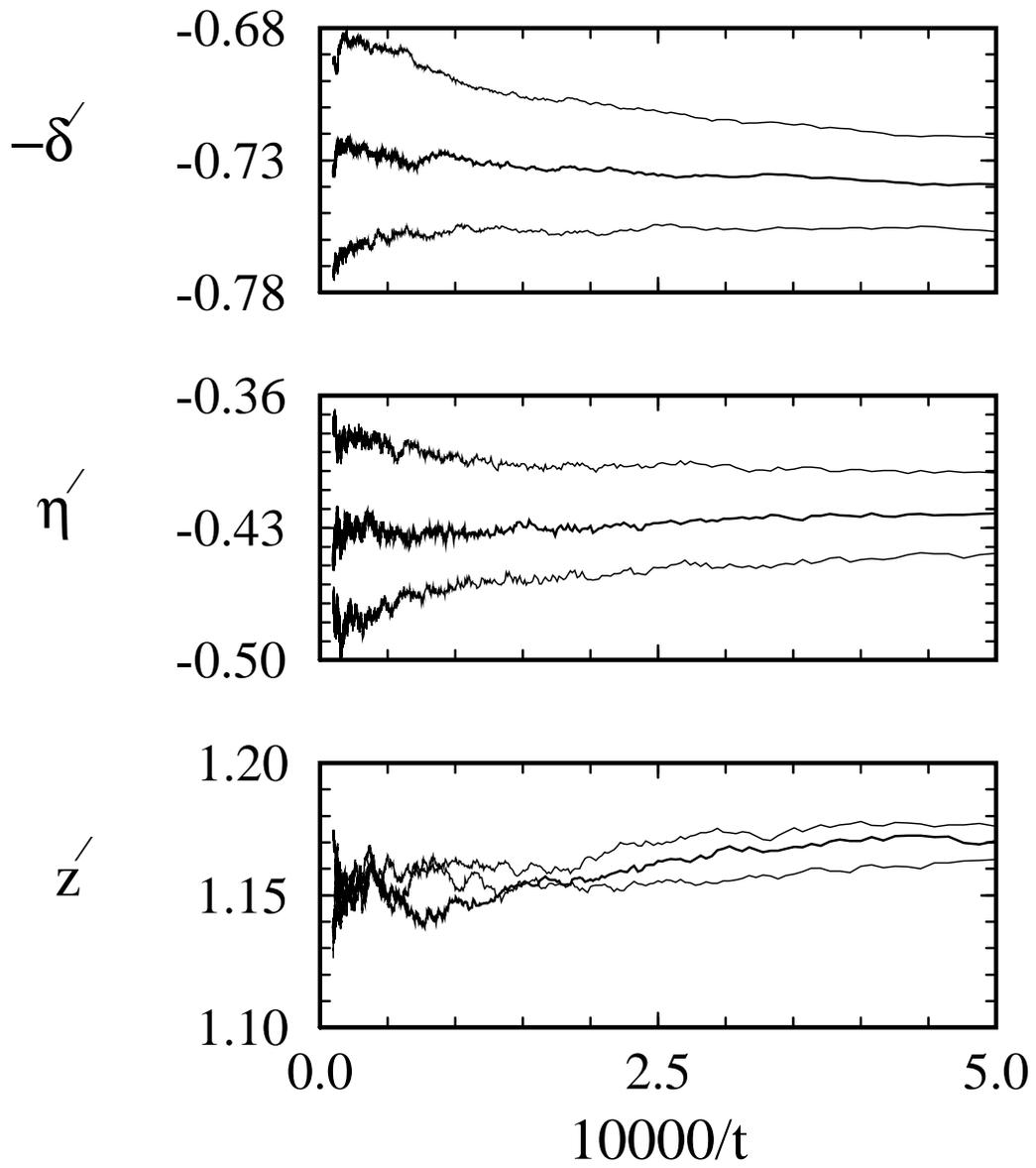,width=6.0in}}
\vspace{0.5in}
\caption{
Effective exponents, as in Fig.~\protect\ref{fig7},
for the second type of interface dynamics
near the bicritical point where the
A and B poisoned phases meet the reactive phase.}
\label{figdyn2}
\end{figure}

In the second type of interface dynamics simulations, which has not
been studied before, the simulation is stopped if the interface between
the domains has ``collapsed'' back to one vacant site.  We introduce a
probability of avoiding a collapse $P(t)\propto t^{-\delta'}$ and
corresponding vacancy concentrations $\langle n(t)\rangle \propto
t^{\eta'}$ and $\langle R^{2}(t)\rangle \propto t^{z'}$.
Figure~\ref{figdyn2} shows results from $10^7$ independent runs each
lasting up to $10^5$ time steps.  We find values of $\delta' =
0.73(2)$, $\eta'=-0.43(2)$ and $z'=1.15(2)$.

Note the value of the dynamic exponent $z$ or $z'$, which measures the
size of the active region during surviving runs, is the same in both
types of interface dynamics simulations as that measured for the defect
dynamics.  Furthermore, although the exponents $\delta$ and $\eta$ are
different in the three cases, their sum $\delta + \eta$ (or
$\delta'+\eta'$), which governs the time evolution of the number of
vacancies in just the surviving runs, are the same within statistical
error.  This indicates a universal nature of the critical spreading of
the active region for models with two symmetric adsorbing states which
is independent of whether defect or interface dynamics is being
considered.  A similar result holds for some one-dimensional systems
with infinitely many adsorbing states\cite{ManyAS}.

Assuming this conjecture is true, it should be noted that simulations
using the first type of interface dynamics, where $\delta \equiv 0$,
yield no information beyond that obtainable from simulations employing
defect dynamics.  However, simulations using the second type of
interface dynamics measure an independent dynamic exponent $\delta'$
which we expect to be a universal number. Recent measurements on
similar models support this conjecture \cite{intheworks1}. 

\subsection{Crossover from bicritical to critical behavior}

\begin{figure}
\vspace{1in}
\centerline{\psfig{file=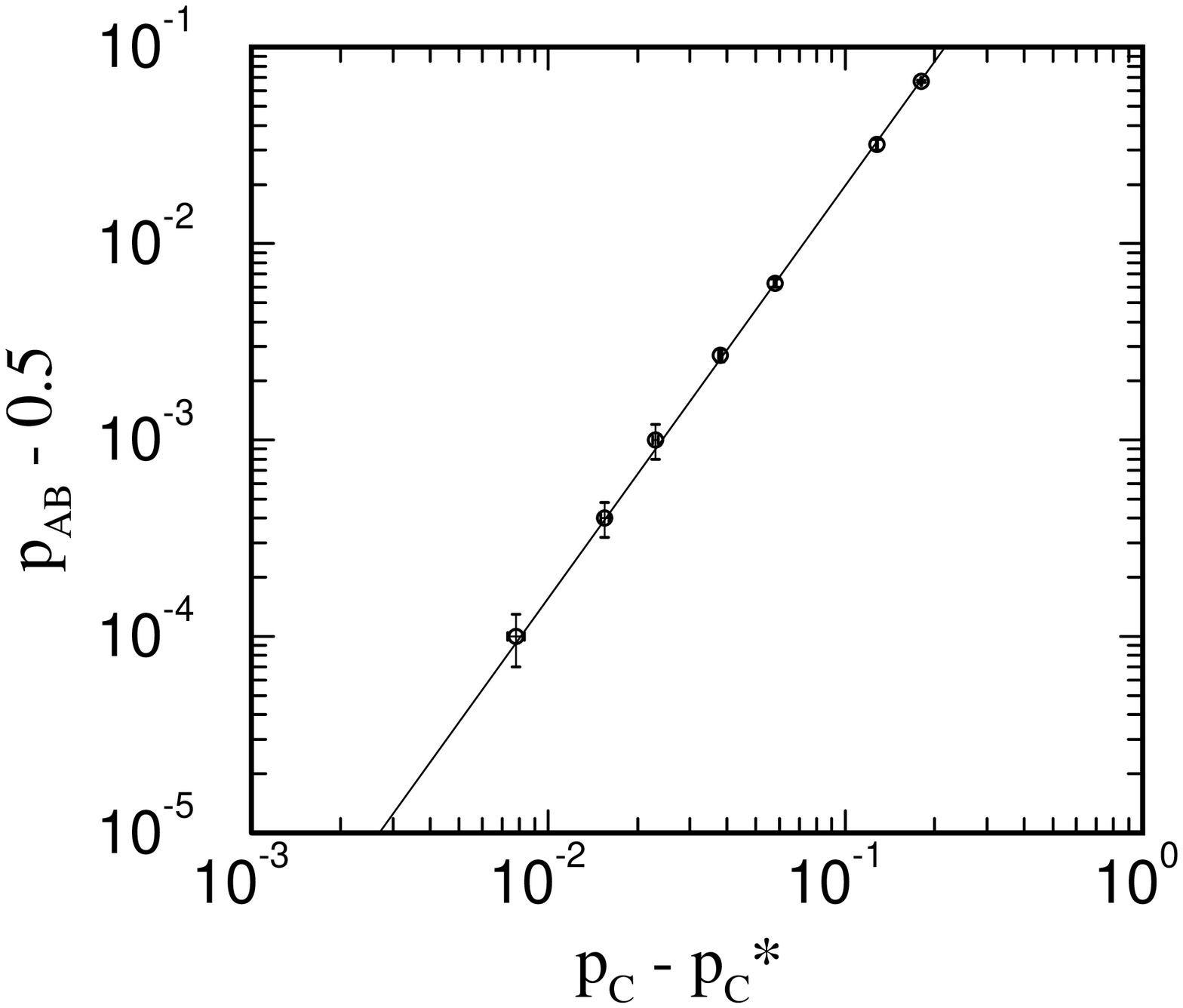,width=6.0in}}
\vspace{0.5in}
\caption{
Location of the critical line as a function of distance from the
bicritical point.  The data falls on a line with a slope
corresponding to the crossover exponent $\phi = 2.1\pm0.1$.}
\label{figcrossexp}
\end{figure}

Finally, we measured the crossover exponent from bicritical to critical
behavior.  Near the bicritical point where the A and B poisoned phases
meet, the boundary of the reactive region is expected to behave as
$(p_{AB} - 0.5) \propto (p_C - p^{*}_C)^{\phi}$, where $\phi$ is the
crossover exponent\cite{BCP}.  We used the dynamical simulation method to
accurately determine the location of the DP phase boundary between the
reactive phase and the $A$ saturated phase near the bicritical point.
From the log-log plot of
$p_{AB}-0.5$ versus $p_C-p^{*}_C$ shown in Fig.~\ref{figcrossexp},
we find $\phi=2.1\pm0.1$.
Our determination of $\phi$ is not as accurate as the other exponents
due to complications arising from crossover effects.
Similar
to the crossover from bicritical to sub-critical behavior described
above, near the bicritical point at short times the dynamical behavior
is controlled by the bicritical point before changing to the directed
percolation critical behavior at long times. The time with which the
crossover occurs increases as the bicritical point is approached,
making studies very close to the bicritical point too time-consuming.

\section{Summary}

We have studied a simple three species monomer-monomer reaction model
to investigate the role of symmetry in adsorbing phase transitions.  We
have shown that, unlike the two species monomer-monomer model or the
monomer-dimer ZGB model, this model has a reactive steady state in one
dimension. There are also poisoned states for which the lattice is
covered by one of the monomer species.  The continuous phase
transitions between the reactive phase and the poisoned states meet at
bicritical points.  Along the first-order coexistance line between two
absorbing phases the model reduces to the two species monomer-monomer
model.

We also constructed a mean-field theory of the model. An unusual
feature of the mean-field analysis is that the bicritical points
lie on the edge of the phase diagram if the correlations 
of triplets of adjacent sites are not exactly treated.
Only when correlations up to triplets of adjacent sites are included does the
bicritical point appear inside the phase diagram, indicating the
importance of reproducing the correlations induced by large domains of
a single saturated phase.

The dynamic critical behavior at the transition between the reactive
phase and a poisoned phase is in the DP universality class. At the
bicritical points, where there are two equivalent poisoned states, the
dynamic critical behavior is in the BAWe class. Thus, the universality
class of the transition changes from DP to BAWe when the symmetry of
the adsorbing state is increased from one to two equivalent noiseless
states.  Furthermore, we have shown that having a two-fold symmetry in the
adsorbing states introduces additional features in the dynamics over
a model with 
a unique adsorbing state.  In particular, the
critical dynamics of the interfaces between two different adsorbing
states shows a sensitivity to how the dynamics is defined, and the
survival probability of  fluctuations in the size of the interface from
its smallest value is described by a new universal exponent $\delta'$.
However, the critical spreading of the reactive region, be it a defect
in a single phase or a domain wall between phases, appears to be
insensitive to the choice of initial conditions.  This appears to
result from the fact that large reactive regions are insensitive to
whether the reactive regions are bounded by the same or different
saturated phases. We do not expect this result to be true in higher
dimensions where the entropy of domain walls can play a role and
nonuniversal critical spreading has been observed in other
models\cite{nonunivspreading}.

This work was supported by the National Science Foundation
under Grant No.~DMR--9408634.

\appendix
\section*{Rate equations in the triplet approximation}

The triplet approximation replaces the actual lattice configuration with
the average configuration of each cluster consisting of a three adjacent
sites. Define the average number of
the different number of triplets as
$$
x_{ijk} \equiv {N_{ijk} \over N}
$$
where $N$ is the total number of triplets, which in one dimension is
equal to the the number of sites, and
$N_{ijk}$ is the number of triplets
consisting of $i$, $j$, and $k$ type monomers
($A$, $B$, or $C$) or vacancies ($V$).
The densities of asymmetric triplets, i.e. $ijk$ type triplets with
$i \neq k$, are by symmetry assumed to be equal, and are added together.

In the adsorption
controlled limit, triplets with adjacent dissimilar monomers,
e.g. A-B-B, A-C-V, \ldots, are forbidden,
leaving 19 allowed types of triplets.
However, the triplet densities must satisfy 4 separate constraints,
which reduce the number of independent triplet densities to 15.
The first of these constraints, similar to the constraints on the
site and bond densities given by Eq.~(\ref{siteconstraint})
and Eq.~(\ref{pairconstraint}) respectively,
merely conserves the total triplet density
$$
\sum_{ijk}x_{ijk} = 1 .
$$
The other three constraints have no analogues in the site or pair
approximations.  Because each particular lattice site contributes
to three different triplets, and the middle and end positions
of the triplets are not symmetric, the total density of $A$ type monomers
occurring in say the left position of the triplets must be equal to the
the total density of $A$ type monomers
occurring in the middle position of the triplets
$$
x_{AVV} + 2x_{AVA} + x_{AVB} + x_{AVC}
= x_{AAV} + 2x_{VAV} ,
$$
and similarly for $B$ and $C$ type monomers.

The equations of motion of the triplet densities can be written as
$$
{d x_{ijk} \over dt} = \sum_{\alpha} \Delta x_{ijk}^{(\alpha)}
$$
where $\alpha$ refers to the event type, and
$\Delta x_{ijk}^{(\alpha)}$ are the triplet density changes with an event
of type $\alpha$. The different types of events were enumerated above in
the discussion of the pair approximation.
The triplet density changes due to $A \downarrow$ and
$A \downarrow \; AB \uparrow$ events are listed in Tables II and III,
respectively, where
$$
y_{A B}
\equiv
x_{BVV} + x_{BVB} + x_{AVB} + {1\over2} x_{BVC}
$$
and $P_{ijk/Xjk}$ is the conditional probability for an $i$ type monomer
or vacancy to occur next to a $jk$ pair.
For example,
$$
P_{VVV/XVV}
=
{x_{VVV} \over x_{VVV} + {1\over2} x_{AVV} + {1\over2} x_{BVV} + {1\over2} x_{CVV} }
$$
and
$$
P_{VAV/XAV}
=
{x_{VAV} \over x_{VAV} + {1\over2} x_{AAV} } .
$$

\narrowtext

Then taking $x_{VVV}$, $x_{AVV}$, $x_{BVV}$, and $x_{CVV}$
to be the dependent triplet densities, the equations of motion of
the independent triplet densities are
\begin{eqnarray}
{d x_{AAA} \over dt}
& = &
p_A \left[ x_{AVA} + \left( 2 x_{AVA} + x_{AVV} \right) \;
P_{AAV/XAV}\right]
\nonumber \\
& &
- \left( p_B \; y_{BA} \;
+ p_C \; y_{CA} \right) \;
P_{AAV/XAV} \; P_{AAA/XAA}
\nonumber
\end{eqnarray}

\widetext

\begin{eqnarray}
{d x_{AAV} \over dt}
& = &
p_A \left[ x_{AVV} + \left( 2 x_{AVA} + x_{AVV} \right) \;
\left( P_{VAV/XAV} - P_{AAV/XAV} \right)
\right]
\nonumber \\
& &
+ \left( p_B \; y_{BA} \;
+ p_C \; y_{C A} \right) \;
P_{AAV/XAV} \;
\left( - 1 + P_{AAA/XAA} - P_{AAV/XAA} \right)
\nonumber
\end{eqnarray}

\begin{eqnarray}
{d x_{VAV} \over dt}
& = &
p_A \left[ x_{VVV} - \left( 2 x_{AVA} + x_{AVV} \right) \;
P_{VAV/XAV} \right]
\nonumber \\
& &
- \left( p_B \; y_{BA}
+ p_C \; y_{CA} \right) \;
\left( P_{VAV/XAV} - P_{AAV/XAV} \; P_{AAV/XAA} \right)
\nonumber
\end{eqnarray}

\begin{eqnarray}
{d x_{AVA} \over dt}
& = &
p_A \left[ - x_{AVA} + \left( 2 x_{VVV} + x_{AVV} \right) \;
P_{AVV/XVV} \right]
\nonumber \\
& &
- \left( p_B + p_C \right) \; x_{AVA}
\nonumber \\
& &
- \left( p_B \; y_{BA}
+ p_C \; y_{CA} \right) \;
P_{VAV/XAV} \; P_{AVA/AVX}
\nonumber
\end{eqnarray}

\begin{eqnarray}
{d x_{AVB} \over dt}
& = &
p_A \; \left( 2 x_{VVV} + x_{AVV} \right) \; P_{BVV/XVV}
\nonumber \\
& &
+ p_B \; \left( 2 x_{VVV} + x_{BVV} \right) \; P_{AVV/XVV}
\nonumber \\
& &
- p_A \; x_{AVB} - p_A \; y_{AB} \;
P_{VBV/XBV} \; P_{AVB/BVX}
\nonumber \\
& &
- p_B \; x_{AVB} - p_B \; y_{BA} \;
P_{VAV/XAV} \; P_{AVB/AVX}
\nonumber \\
& &
- p_C \; x_{AVB}
\nonumber \\
& &
- p_C \; y_{CA} \;
P_{VAV/XAV} \; P_{AVB/AVX}
- p_C \; y_{CB} \;
P_{VBV/XBV} \; P_{AVB/BVX}
\nonumber
\end{eqnarray}
and similarly for the remaining densities.

\narrowtext

\begin{table}

\begin{tabular}{lcc}
event type & $A\downarrow$ & $A\downarrow \; AB\uparrow$ \\ \hline
$\Delta x_{VV}$ &
$-2 \; p_A  \; z_{AV} \; x_{VV}$ &
${1\over2} p_A \; x_{BV} \; (1+z_{AV}) \left[ 1+ {x_{BV} \over 2x_{B}}\right]$
\\
$\Delta x_{AA}$ &
$p_A \; z_{AV} \; x_{AV}$ &
$0$
\\
$\Delta x_{BB}$ &
$0$ &
$-{1\over2} p_A \; x_{BV} ( 1 + z_{AV}) {x_{BB} \over x_{B}}$
\\
$\Delta x_{CC}$ &
$0$ &
$0$
\\
$\Delta x_{AV}$
& $p_A \; z_{AV}(2x_{VV}-x_{AV})$
& $0$
\\
$\Delta x_{BV}$
& $0$
& $-{1\over2} p_A \; x_{BV} (1 + z_{AV}) {x_{BV}\over x_B}$
\\
$\Delta x_{CV}$
& $0$
& $0$
\\
\end{tabular}

\caption{Bond density changes for different events in the pair approximation.}

\end{table}

\begin{table}
\begin{tabular}{lc}
event type
& $A\downarrow$
\\ \hline
$\Delta x_{VVV}$
& $- p_A \left[ x_{VVV} + ( 2 x_{VVV} + x_{AVV} ) P_{VVV/XVV} \right]$
\\
$\Delta x_{AAA}$
& $p_A \left[ x_{AVA} + ( 2 x_{AVA} + x_{AVV} ) P_{AAV/XAV}\right]$
\\
$\Delta x_{AAV}$
& $p_A \left[x_{AVV} + (2 x_{AVA} + x_{AVV}) (P_{VAV/XAV} - P_{AAV/XAV})
\right]$
\\
$\Delta x_{VAV}$
& $p_A \left[ x_{VVV} - ( 2 x_{AVA} + x_{AVV} ) P_{VAV/XAV} \right] $
\\
$\Delta x_{AVA}$
& $ p_A \left[ - x_{AVA} + ( 2 x_{VVV} + x_{AVV} ) P_{AVV/XVV} \right]$
\\
$\Delta x_{AVV}$
& $p_A \;
\left[ -x_{AVV} + (2 x_{VVV} + x_{AVV}) (P_{VVV/XVV} - P_{AVV/XVV}) \right] $
\\
$\Delta x_{BVV}$
& $- p_A (2 x_{VVV} + x_{AVV}) P_{BVV/XVV}$
\\
$\Delta x_{CVV}$
& $- p_A (2 x_{VVV} + x_{AVV}) P_{CVV/XVV}$
\\
$\Delta x_{AVB}$
& $p_A (2 x_{VVV} + x_{AVV}) P_{BVV/XVV}$
\\
$\Delta x_{AVC}$
& $p_A (2 x_{VVV} + x_{AVV})  P_{CVV/XVV}$
\\
\end{tabular}
\caption{Triplet density changes due to an $A$ monomer adsorbing,
and remaining on the lattice, in the triplet approximation.}
\end{table}

\begin{table}
\begin{tabular}{lc}
event type
& $A\downarrow \; AB \uparrow$
\\ \hline
$\Delta x_{VVV}$
& $p_A \; x_{BVV}
+ p_A \; y_{AB} \;
P_{VBV/XBV} ( 1 + P_{BVV/BVX} ) $
\\
$\Delta x_{BBB}$
&
$ - p_A \; y_{AB} \;
P_{BBV/XBV} \; P_{BBB/XBB} $
\\
$\Delta x_{BBV}$
&  $ p_A \; y_{AB} \;
P_{BBV/XBV} \;
\left( - 1 + P_{BBB/XBB} - P_{BBV/XBB} \right)$
\\
$\Delta x_{VBV}$
& $ - p_A \; y_{AB} \;
\left( P_{VBV/XBV} - P_{BBV/XBV} \; P_{BBV/XBB} \right) $
\\
$\Delta x_{BVB}$
& $- p_A \; x_{BVB} - p_A \; y_{AB} \;
P_{VBV/XBV} \; P_{BVB/BVX}$
\\
$\Delta x_{AVV}$
& $p_A \; x_{AVB} + p_A \; y_{AB} \;
P_{VBV/XBV} \; P_{AVB/BVX} $
\\
$\Delta x_{BVV}$
& $p_A \left[ x_{BVB} - x_{BVV}  \right]
+ p_A \; y_{AB} \;
P_{BBV/XBV} $
\\
&
$ + p_A \; y_{AB} \;
P_{VBV/XBV} \; \left(  P_{BVB/BVX} - P_{BVV/BVX} \right) $
\\
$\Delta x_{CVV}$
& $p_A \; {1\over2} x_{BVC} + p_A \; y_{AB} \;
P_{VBV/XBV} \; P_{BVC/BVX}$
\\
$\Delta x_{AVB}$
& $- p_A \; x_{AVB} - p_A \; y_{AB} \;
P_{VBV/XBV} \; P_{AVB/BVX}$
\\
$\Delta x_{BVC}$
& $- p_A \; {1\over2} x_{BVC}
- p_A \; y_{AB} \;
P_{VBV/XBV} \; P_{BVC/BVX}$
\\
\end{tabular}
\caption{Nonvanishing 
triplet density changes due to an $A$ monomer adsorbing,
and reacting with a $B$ monomer to form a $AB$ molecule,
in the triplet approximation.}
\end{table}

\begin{references}
\bibitem{ctsnote}
Actually, this statement is known to be true only
for directed percolation \protect\cite{DP2}, but it seems to also be
true\protect\cite{BAWe} for the BAWe universality class.
\end{references}
\end{document}